\documentclass[a4paper]{jpconf}
\usepackage{graphicx}

\newcommand\Schr{Schr\"odinger}
\newcommand\sqtwoinv{\frac{1}{\sqrt{2}}}

\begin{document}
\title{Nonlinear \Schr~equation in foundations: summary of 4 catches}

\author{Lajos Di\'osi}

\address{Wigner Research Centre for Physics, Budapest 114, P.O.Box 49, H-1525 Hungary}

\ead{diosi.lajos@wigner.mta.hu}

\begin{abstract}
Fundamental modifications of the standard \Schr~equation by
additional nonlinear terms have been considered for various purposes
over the recent decades. It came as a surprise when, inverting Abner
Shimonyi's observation of "peaceful coexistence" between standard
quantum mechanics and relativity, N. Gisin proved in 1990 that any
(deterministic) nonlinear \Schr~equation would allow for superluminal
communication. This is by now the most spectacular and best known
anomaly. We discuss further anomalies, simple
but foundational, less spectacular but not less dramatic.
\end{abstract}

\section{Introduction}
We discuss major difficulties encountered by nonlinear modifications
of the \Schr~equation. Before we enter the subject, we make it clear
that nonlinear \Schr~equations and nonlinear effective modifications
of quantum mechanics are widely used  in physics as approximate means. 
Most typical is the mean-field approximation.  It yields the nonlinear Hartree-Fock
equation  \cite{Hartree} so basic for many-electron systems, and it yields the semi-classical 
Einstein equation \cite{scEinstein} which is indispensable in quantum cosmology. 
These are approximate (effective) nonlinear quantum mechanics.
One can, on the contrary, consider nonlinear modification of quantum mechanics
at the fundamental level. As early as in 1952, J\'anossy proposed \cite{Janossy}
that a simple state-dependent potential ensure localized stationary wave
function of  a free quantum particle (of mass M):
\begin{equation}\label{Janossy}
i\hbar\frac{d\Psi(x)}{dt}=-\frac{\hbar^2}{2M}\Psi^{\prime\prime}(x)
                                         +\frac{1}{2}\alpha^2(x-\langle x \rangle_\Psi)^2\Psi(x),
\end{equation}  
where 
\begin{equation}
\langle x \rangle_\Psi=\int x\vert\Psi(x)\vert^2 dx
\end{equation}
and $\alpha$ is a certain parameter. The new contractive `mean-field'
potential counters the unlimited dispersion predicted by the standard
\Schr~equation. The stationary states are localized Gaussians (solitons).
J\'anossy's equation was and remained 
forgotten until now \cite{Gao}.  A gravity-related version of his nonlinear
term was suggested much later independently by the author and by
Penrose \cite{Dio84,Pen}, both being unaware of J\'anossy's equation yet seeking the same effect, 
namely,  to ensure localization of quantum objects --- massive ones this time.
We choose this equation, called the \Schr---Newton equation (SNE), to be our testbed
to discusse four typical catches otherwise valid for any nonlinear 
\Schr~equations.

 \section{Peaceful coexistence}
It has been known since the beginning of quantum theory that it 
contains a certain latent non-locality. For instance,  
the collapse of the single particle wave function, occurring in quantum measurement, is 
to happen instantaneously over whatever large distances covered by the
pre-measurement wave function.  For two distant particles, Einstein, Podolski and Rosen (EPR) in \hfill\break
\begin{minipage}[b]{0.84\textwidth}
1935 \cite{EPR} constructed the sharpest example of apparent action-at-a-distance
and John Bell in 1963 \cite{Bell} pointed out a specific non-locality. 
It was also clear that  neither the apparent action-at-a-distance nor the quantum non-locality
could result in testable dynamical effects like dynamical action-at-a-distance or superluminal communication.
The physicist and philosopher Abner Shimony formulated this paradoxical situation as
the peaceful coexistence between quantum theory and special relativity \cite{Shimony}.
\end{minipage}
\begin{minipage}[t]{0.15\textwidth}
\includegraphics[width=\textwidth]{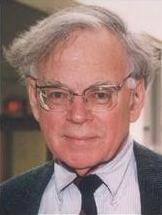}
\end{minipage}

The reason of why formal violations of locality will all cancel from the measurable
outcomes of quantum mechanics lie in the linear structure of the mathematical
representation of the state space, dynamics, and measurable outcomes themselves.
If one adds non-linear terms to the \Schr~equation the peaceful coexistence
ends and the physical violation of special relativity takes over. Interestingly,
J\'anossy \cite{Janossy} mentions that his non-linear
single particle \Schr~equation suffers of superluminal physical influence.
The most spectacular thesis belongs to Gisin \cite{Gisin}.
Any nonlinear modification of the single particle \Schr~equation,
\begin{equation}\label{NLSE}
i\hbar\frac{d\Psi(x)}{dt}=-\frac{\hbar^2}{2M}\Psi^{\prime\prime}(x)
                                                +V_\Psi(x)\Psi(x),
\end{equation}
allows for superluminal communication between the  EPR partners for whatever small 
state-dependent potential $V_\Psi(x)$ (apart from the constant potential, of course).

\section{\Schr---Newton equation: our testbed}
Consider the single-body SNE for the centre-of-mass free motion of a `large' composite 
object of mass $M$:
\begin{equation}\label{SNE}
i\hbar\frac{d\Psi(x)}{dt}=-\frac{\hbar^2}{2M}\Psi^{\prime\prime}(x)
                                                +M\Phi_\Psi\Psi(x),
\end{equation}
where the state-dependent Newton potential reads
\begin{equation}\label{mean-field}
\Phi_\Psi(x)=-GM\int\frac{\vert\Psi(r)\vert^2}{\vert x-r\vert}d^3r.
\end{equation}
Although this equation is the Newtonian limit of  the approximate semiclassical Eisntein
equation, it was suggested that it might be
fundamental \cite{Dio84,Pen} and it is currently studied as such, 
both theoretically and experimentally \cite{SNEs}.

Our forthcoming analysis needs the following peculiar features of the SNE.
It possesses standing (static) soliton solutions $\vert\bigcirc\rangle$ of diameter
\begin{equation}\label{D}
D\sim\frac{\hbar^2}{GM^3},
\end{equation}
c.f., e.g., \cite{Dio84}.
We can construct the \Schr~cat states formed by two separate static solitons: 
\begin{equation}\label{Cat}
\Psi_{\pm}=\sqtwoinv\left(\vert\bigcirc_L\rangle\pm\vert\bigcirc_R\rangle\right).
\end{equation}
The left and rigth solitons $\bigcirc_{L/R}$ are seperated by a distance $\ell\gg D$ initially.
Because of the mean-field $\Phi_\Psi(x)$ in the SNE, the two solitons in $\Psi_\pm$ attract each other 
and start to move, like this: 
$$
\bigcirc\!\!\!\!\!\!\longleftarrow\!\!\ell\!\!\longrightarrow\!\!\!\!\!\!\bigcirc
~~\Longrightarrow~~
\bigcirc\bigcirc
~~\Longrightarrow~~
\bigcirc\!\!\!\!\!\bigcirc
~~\Longrightarrow~~
\bigcirc\bigcirc
~~\Longrightarrow~~
\bigcirc\!\!\!\!\!\!\longleftarrow\!\!\ell\!\!\longrightarrow\!\!\!\!\!\!\bigcirc
~~\Longrightarrow~~
\bigcirc\bigcirc
~~\Longrightarrow~~\dots
$$
They are oscillating as if they were two (interpenetrating) gravitating bodies of mass $M/2$ each, 
the periodicity is $T=\pi\sqrt{\ell^3/2GM}$ known from two-body problem.

The feature we should note is the following. 
Single soliton states $\vert\bigcirc_L\rangle$ and $\vert\bigcirc_R\rangle$ are static.
The two-soliton \Schr~cat states $\Psi_\pm$  oscillate. 
The initial overlap of $\Psi_\pm$ with $\bigcirc_{L/R}$ is  $1/\sqrt{2}$, see (\ref{Cat}).  In standard
linear quantum dynamics  the overlap would be constant in time. It is not so here,
the overlap will oscillate with amplitude $1/\sqrt{2}$. In fact, it is zero for most
of the time.  The SNE makes $\Psi_\pm(t)$ orthogonal to  $\vert\bigcirc_{L/R}\rangle$ after a few times
\begin{equation}\label{t}
\delta t\sim\frac{\hbar\ell}{GM^2}\ll T,
\end{equation}
when the distance differs from $\ell$ by (a few times) more than the soliton size $D$. 
To confirm the guess, consider the relative acceleration $GM\ell^{-2}$ of the
two solitons toward each other  and calculate the time lapse until  they get closer by a length $\sim\!\!D$.

\section{\Schr---Newton equation: four catches}
Based on the above features of \Schr~cat states under SNE, we are going to
present four interrelated foundational issues all originating exclusively from the 
nonlinearity. Our thought experiments adapt Gisin's two-qubit superluminal telegraph  \cite{Gisin}.
Suppose Alice and Bob are far away from each other. 
Alice owns a qubit and  Bob owns a large mass $M$, in the 
following maximally entangled composite state:
\begin{equation}\label{entangled}
   \vert\uparrow_z\rangle\otimes\vert\bigcirc_L\rangle
 +\vert\downarrow_z\rangle\otimes\vert\bigcirc_R\rangle.
\end{equation}
Three of the forthcoming testable controversies are based on this composite state,
the fourth one uses the single particle dynamics in itself. 

\subsection{Action-at-a-distance from nothing}\label{AAD}
Alice measures either $\hat\sigma_z$ (case i) or $\hat\sigma_x$ (case ii), each
with random outcomes $\pm1$. 
In case i, Bob's state collapses into the static single soliton states $\vert\bigcirc_L\rangle$ or $\vert\bigcirc_R\rangle$, 
according to the outcomes $\pm1$.  Alternatively, in case ii, 
Bob's state collapses into the \Schr~cat states $\Psi_\pm$,
respectively. So far the story coincides with the
standard EPR scenario of standard quantum mechanics. 
The salient effect of nonlinearity enters from now on. 
In case i, Bob's states $\vert\bigcirc_L\rangle$ or $\vert\bigcirc_R\rangle$ remain
static. In case ii, his states $\Psi_\pm$ heavily oscillate.  
Using no physical interaction with it,  Alice could cause a testable effect to the remote object of Bob.  

\subsection{Superluminal telegraph}\label{FTL}
This thought experiment is the continuation of the previous one.
Bob tests whether his initial state is preserved (case i) or
it is changing (case ii). Bob waits until a few times $\delta t$ (\ref{t}) and then he measures the projector
\begin{equation}\label{projector}
\vert\bigcirc_L\rangle\langle\bigcirc_L\vert+\vert\bigcirc_R\rangle\langle\bigcirc_R\vert.
\end{equation}
The outcome is $1$ in case i and $0$ in case ii.
This confirms that the action-at-a-distance (Sec.~\ref{AAD}) is testable. 
Is it superluminal? 
It is, if the distance between Alice and Bob exceeds $c\delta t$. 

\subsection{Unsuitability for  mixed state}
The initial setup is the same as before, Alice and Bob are far away from each other
but they are not assumed to cooperate this time. They may even not know about
each other yet they are supposed to inherit the entangled states (\ref{entangled}).
We assume that Alice does not measure her qubit at all or, if she measures anything,
Bob cannot learn anything about the measurement and the outcome. 
Bob's local (reduced) state is anyway a  mixed state 
\begin{equation}
\rho=\frac{1}{2}
\Bigl(|\bigcirc_L\rangle\langle\bigcirc_L|+                            
          |\bigcirc_R\rangle\langle\bigcirc_R|
\Bigr).
\end{equation}
The SNE does, as a matter of fact, not apply to mixed states but pure ones
described by a wave function. It is clear now that Bob cannot calculate the
dynamical evolution of his local system. 
This incapacity is general since under natural conditions local quantum systems
are never in pure states since they are always entangled with the rest of the world or with their environment at least.

\subsection{Breakdown of statistical interpretation}
To understand this catch, we need no EPR situation but the single-body SNE in itself. 
Consider that the SNE evolves a given initial pure state density matrix $\rho^i$ 
into a given final pure state density matrix  $\rho^f$ and note that the map
\begin{equation}
\rho^f = \mathcal{M}[\rho^i]
\end{equation}
is nonlinear.

We are going to show that any nonlinear map makes the statistical interpretation
of quantum theory impossible. The proof is elementary and quick \cite{DiosiQI}.
Consider the weighted statistical mixing of two states $\rho_1,\rho_2$ with weights 
$\lambda_1+\lambda_2=1$, the resulting state reads
\begin{equation}
\rho=\lambda_1\rho_1+\lambda_2\rho_2.
\end{equation}
In von Neumann standard theory, the order of mixing and dynamical evolution are interchangeable:
\begin{equation}
\mathcal{M}[\lambda_1\rho_1+\lambda_2\rho_2]=
\lambda_1\mathcal{M}[\rho_1]+\lambda_2\mathcal{M}[\rho_2].
\end{equation}
Now we recognize that this is the mathematical condition of $\mathcal{M}$'s linearity!
From the structure of the proof we see that the interchangeability of mixing and
dynamical map excludes any  nonlinear  \Schr~equation not just the SNE.
Without such interchangeability the statistical interpretation of quantum states
totally collapses. Moreover, the principle of interchangeability of mixing and dynamics
is not necessarily quantum, it is carved in marble in classical statistical physics. 

\section{Summary: catches and loopholes}
We listed four foundational issues encountered by nonlinear
modifications of the \Schr~equation. Our example was
the so-called \Schr---Newton equation but all four anomalies are equally
valid for any deterministic nonlinear modification of the dynamics of any 
simple or composite quantum system. The best known and most spectacular
anomaly is superluminality, a clear violation of special relativity. The derivation
of superluminality exploits nonlinearity together with standard statistical interpretation
of the wave function. Less known is that already the statistical interpretation 
is in ultimate conflict with any (deterministic) nonlinearity of quantum dynamics, 
cf., e.g., \cite{Mielnik,Diosi86}. 
This anomaly is admittedly less spectacular than superluminality. 
It is, nonetheless, the deepest anomaly of nonlinear quantum mechanics.

Non-linear quantum mechanics,  the SNE first of all, attracts growing attention in foundations. 
Awareness of the also foundational catches has recently
motivated a stochastically re-linearized model \cite{NimHor} and theory \cite{TilDio}.
The surviving and even growing interest in SNE as it is can be 
explained and perhaps justified despite the catches if we point out certain
loopholes.  In summary, nonlinear \Schr~equations 
\begin{itemize}
\item{allow for}  
    \begin{itemize}
    \item{fake action-at-a-distance but it may be extreme weak to be detected}
   \item{superluminal communication but it may be too hard to be realized}
   \end{itemize}
\item{does not allow for}
   \begin{itemize}
   \item{local dynamics unless a local pure state is prepared}
   \item{standard statistical interpretation hence a new interpretation is needed}
   \end{itemize}
\end{itemize}
The last catch is the major one. {\it Any} nonlinear quantum theory needs a radical new
interpretation of the wave function because {\it any} (deterministic) nonlinear dynamics makes 
Born interpretation contradict to principles of statistics. 

\ack
This work was partially supported by the
Hungarian Scientific Research Fund under Grant No. 103917 and
EU COST Action MP1209 `Thermodynamics in the quantum regime'. 
The author is indebted to Franklin Fetzer Fund for its continued 
generous support of foundational research.

\end{document}